\documentclass[a4paper]{jpconf}
\usepackage{graphicx}
\usepackage{mathrsfs}
\usepackage{amssymb}
\usepackage{amsmath}
\usepackage{graphicx}
\usepackage{bm}
\begin{document}
\title{Directional Detection of Dark Matter with MIMAC}

\author{J.~Billard, F.~Mayet and D.~Santos}

\address{Laboratoire de Physique Subatomique et de Cosmologie, Universit\'e Joseph Fourier Grenoble 1,
  CNRS/IN2P3, Institut Polytechnique de Grenoble, Grenoble, France}

\ead{billard@lpsc.in2p3.fr}

\begin{abstract}
 Directional detection is a promising search strategy to discover galactic Dark Matter.  
We present a Bayesian analysis framework dedicated to Dark Matter phenomenology using directional detection. 
The interest of directional detection as a powerful tool to set exclusion limits, to authentify a Dark Matter detection or to constrain the 
Dark Matter properties, both from
particle physics and galactic halo physics, will be demonstrated.
 However, such results need highly accurate track reconstruction which should be reachable by the MIMAC detector
using a dedicated readout combined with a likelihood analysis of recoiling nuclei.
\end{abstract}

\section{Introduction}
Taking advantage of the astrophysical framework, directional detection of Dark Matter is an interesting strategy  to distinguish
 WIMP events from background ones \cite{spergel}.
Indeed, like most spiral galaxies, the Milky Way is supposed to be immersed in a halo of WIMPs which outweighs the luminous component by at 
least one order of magnitude. As the Solar System rotates around the galactic center through this Dark Matter halo, WIMP events should mainly come
 from the direction to which points the
Sun velocity vector and which happens to be roughly in the direction of the constellation Cygnus ($\ell_\odot = 90^\circ,  b_\odot =  0^\circ$).
  Hence, we argue that a clear and unambigous signature of a Dark Matter
  detection could be achieved by showing the correlation of the measured signal with the direction of the solar motion.\\
  However, the simultaneous measurement of the recoil energy and the recoil direction of a nucleus target of the detector material 
  is a major experimental challenge. In the case of a Fluorine
target at 50 mbar, the mean track length of a 10 keV recoil is of the order of 300 $\mu m$ and about 3 mm at 100 keV. Then, directional detection of Dark Matter
requires very sensitive experiment combined with highly performant technology. In this context, the MIMAC project has been proposed \cite{mimac} and the track reconstruction
strategy and some expected resolutions will presented hereafter.
 
 \section{Directional Detection Phenomenology}

\begin{figure}[t]
\begin{center}
\includegraphics[scale=0.35]{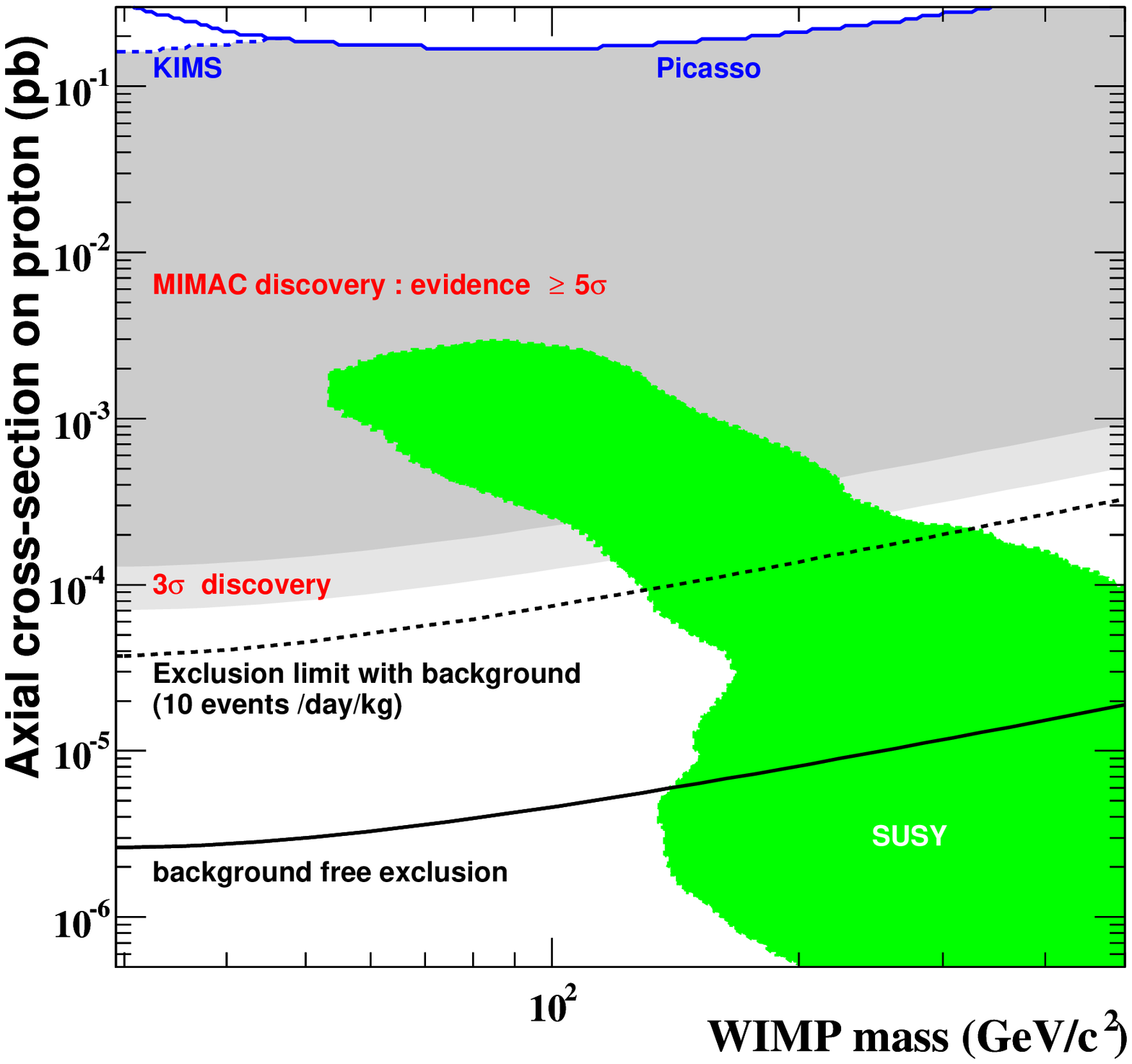}
\includegraphics[scale=0.40,angle=0]{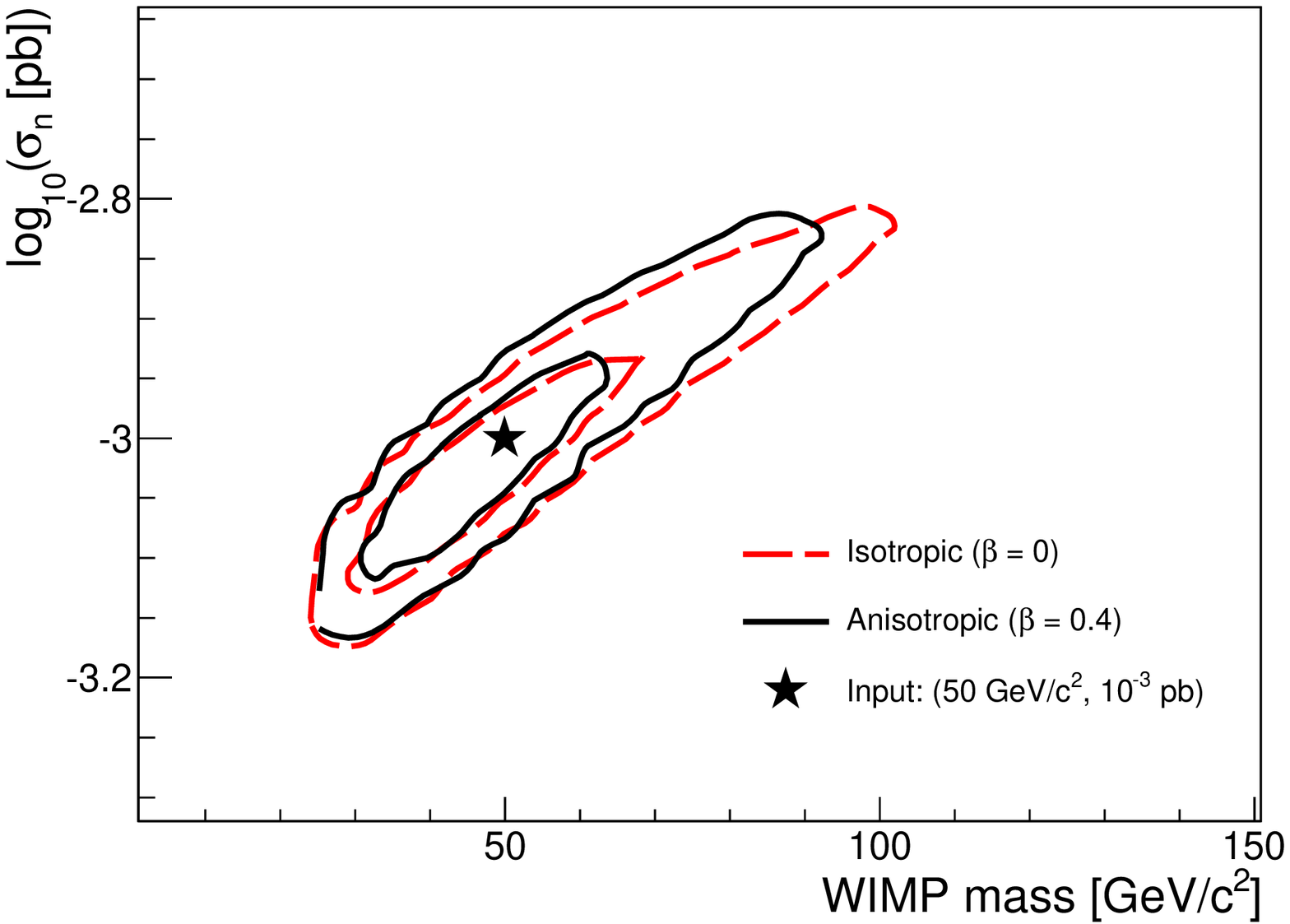}
\caption{Left: Spin dependent cross-section on proton (pb) as a function of the WIMP mass ($\rm GeV/c^2$). 
Exclusion limits from some direct detection experiments are presented, KIMS~\cite{kims} and Picasso~\cite{picasso} as well as
the theoretical region, obtained within the framework of the constrained minimal supersymmetric model from \cite{superbayes}.
Contours corresponding to a significance greater than 3$\sigma$ and 5$\sigma$ are presented in dark and light grey.
The exclusion limit corresponding to pure background data is presented as the black dashed line and the detector sensitivity as the black solid line. Right: 
68\% and 95\% contour level in the ($m_{\chi},\sigma_n$) plan, for a 50 $\rm GeV/c^2$ WIMP and for two input models : isotropic ($\beta=0$) and 
triaxial ($\beta=0.4$), and considering a background event rate of 10 events/kg/year.}
\label{fig:discovery}
\end{center}
\end{figure}

  The first step when analysing directional data should be to look for a signal pointing toward the
   constellation Cygnus with a sufficiently high significance \cite{billard.disco}. If no evidence in favor of a Galactic origin of the signal is deduced from the previous
 analysis, then an exclusion limit should be derived \cite{billard.exclusion}.
On the left panel of figure \ref{fig:discovery} we present the projected discovery regions and exclusion limits for a forthcoming directional detector proposed by the MIMAC
 collaboration. We consider 
a  10 kg $\rm CF_4$ detector 
operated during  $\sim 3$ years,  allowing 3D 
track reconstruction, with a $10^\circ$ angular resolution, a  recoil energy range of 5-50 keV and with a conservative background rate of 10 evts/kg/year.
From the left panel of figure \ref{fig:discovery}, two different scenarii may be distinguished:
\begin{itemize}
\item The dark and light-grey shaded areas represent the contours where a directional
detection of Dark Matter would have an expected significance greater than 3$\sigma$ and 5$\sigma$. These contours are deduced from the map-based likelihood method presented in 
\cite{billard.disco}.
As an illustration, if the WIMP-nucleon cross-section is about $10^{-4}$ pb with a WIMP mass of 100 GeV.c$^{-2}$, the detector would have a
 Dark Matter detection with a significance greater than 3$\sigma$. 
\item If the WIMP-nucleon cross-section is lower than $10^{-5}$ pb, then an exclusion limit is deduced using the extended likelihood method (black dashed line) presented in
 \cite{billard.exclusion}. As a benchmark
and to illustrate the effect of background on exclusion limits, the detector sensitivity (no event) is presented (black solid line).
\end{itemize} 
Left panel of figure \ref{fig:discovery} also presents exclusion limits from direct detection experiments, KIMS~\cite{kims} and Picasso~\cite{picasso} as well as
the theoretical region, obtained within the framework of the constrained minimal supersymmetric model taken from \cite{superbayes}. We can conclude that a directional
detector like MIMAC will cover an important region of interest worth being investigated.\\

As shown in \cite{billard.ident}, it is possible to go further by exploiting all the information
 from a directional detector,
{\it i.e.} the energy and the direction of each event. We will then consider the most optimistic scenario where the WIMP nucleon cross-section is sufficiently large
($\sigma_p = 10^{-3}$ pb)
to get a
high significance Dark Matter detection. In such case, we have shown the possibility to 
constrain the  WIMP properties, 
both from particle physics ($m_\chi, \sigma_n$) and galactic Dark Matter halo physics (velocity dispersions) \cite{billard.ident} with a single directional experiment.
To do so, we used a Markov Chain Monte Carlo algorithm dedicated to the sampling of the likelihood function characterized by 8 free parameters which are 
$\{m_{\chi}, \log_{10}(\sigma_n), l_{\odot}, b_{\odot},\sigma_{x}, \sigma_{y}, \sigma_{z}, R_b\}$, 
where the direction $(l_{\odot}, b_{\odot})$ refers to 
the main direction of the recorded events, $\sigma_p$ is the WIMP-nucleon cross section, $m_{\chi}$ the WIMP mass, $\sigma_{x,y,z}$ the velocity dispersions of the
WIMP velocity distribution and $R_b$ is the background event rate. As a result, one can see from the right panel of figure \ref{fig:discovery} that strong and 
consistent constraints on the $(m_{\chi}, \log_{10}(\sigma_n))$ plane can be obtained within the framework of a multivariate Gaussian WIMP velocity distribution, where the
dispersions, and hence the anisotropy, are set as free parameters. Indeed, even in the case of an extremely anisotropic halo model, the WIMP mass and cross section are
consistent with the input values as the velocity distribution is being constrained simultaneously. 
This is the first step toward a Dark Matter halo model independent analysis that could be done using data
from upcoming directional experiment.

\section{Track reconstruction with MIMAC}
\subsection{MIMAC prototype}

The MIMAC prototype is the elementary chamber of a future large matrix of chambers. It allows us to perform a track measurement with the required performance 
to achieve directional detection of Dark Matter.
The primary electron-ion pairs produced by a nuclear recoil in one chamber of the matrix are detected by drifting the electrons to the grid of a bulk micromegas \cite{bulk}
 and by producing the avalanche in a very thin gap (128 or 256$\mu$m). 
 The electrons move towards the grid in the drift space and are projected on the anode thus allowing to get 
information on X and Y coordinates.
To access the X and Y dimensions, a bulk micromegas  with a 10 by 10 cm$^2$ active area segmented in pixels with a pitch of 424 $\mu$m
 is used as a 2D readout.
 In order to reconstruct the third dimension Z of the recoil, a self-triggered electronics has been developed. It allows
  to perform the anode sampling at a frequency of 50 MHz, leading to time slices of 20 ns width along the Z dimension.
This electronic also includes a dedicated 16 channels ASIC \cite{richer} associated to a DAQ \cite{bourrion}. 
The total recoil energy is deduced from the measured ionization quenching factor (IQF) \cite{mimac}.\\

\subsection{Likelihood approach for track reconstruction}

A straightforward track reconstruction algorithm consists in using a three dimensional linear fit of the center of gravity of each time slice. 
However, we found that this method
failed at recovering the correct $\theta$ angle due to the longitudinal diffusion of the electrons. In order to recover correctly the track properties, two strategies can be
used:
\begin{itemize}
\item If one is able to get the Z position of the track in the detector, {\it e.g.} by using a PMT sensitive to the primary scintillation, the electron diffusion can be
correctly estimated and substracted
\item If no additional readout is used, the only way to recover without bias the track properties, is to fit all the parameters at the same time using a likelihood approach.
This is the strategy that we have developed and which is presented hereafter.
\end{itemize}

To perform this likelihood approach dedicated to track reconstruction within the MIMAC experiment, we combined the use of three different softwares: SRIM \cite{srim} for the
track simulation, MagBoltz \cite{Magb} for the estimation of the electron drift properties (velocity and diffusion), and the MIMAC DAQ simulation software. This way, we
are able to simulate tracks as measured by the detector. Of course, this highlights the need for the detector commissioning as any discrepancy between the simulation and the
measurement will result in bias in the track reconstruction. \\

In order to estimate the expected performance of the detector, in terms of spatial and angular resolution, systematical studies have been done. To do so, we have simulated
about one thousand tracks generated isotropically in the lower half sphere (going downward) at a fixed Z coordinate and energy.
In the case of a Fluorine recoil in our gas mixture 
(70\% CF$_4$ + 30\%CHF$_3$), we found that $\sigma_{x,y}$ is about 1 mm at 20 keV and 0.4 mm at 100 keV and that it is independent from the Z coordinate. 
About $\sigma_z$, we found a resolution of about 1.5 mm at 40 keV and 1 mm at 100 keV, with a weak dependence with the Z coordinate.
 The fact that the spatial resolutions
are of the order of a millimeter will allow us to perform an accurate three dimensional fiducialization of the detector volume to prevent from surface events.
The expected angular resolution $\sigma_{\gamma}$ depends strongly 
on the energy of the recoiling nucleus. Indeed,
the angular resolution is about 55$^{\circ}$ at 20 keV while it is about 30$^{\circ}$ at 100 keV. However, it does not depend on the Z coordinate of the
track in the 3 cm to 7 cm range. \\


We used a Boosted Decision Tree \cite{tmva} as a sense recognition discriminant by first finding the maximum Likelihood according to the two hypothesis: Up and Down.
 Then, we generate a large sample of tracks (around 10,000) corresponding to the best fits of the two hypothesis. If the track is
  going upward or downward, the BDT value should be positive or negative respectively.  
  As a working example, we
  have simulated 1000 tracks generated at 100 keV, in the direction $(\theta =
  45^{\circ},\phi=0^{\circ})$, going downward and at 1.3 cm of the anode. We found that 77\% of the tracks 
  have a negative  BDT value, thus corresponding to a 77\% sense recognition efficiency for this given set of parameters (position, direction and energy).

\section{Conclusion}

In this paper, we have shown that directional detection of Dark Matter could be sensitive to very low cross-sections $\sim 10^{-5}$ pb, in the case of a 10 kg CF$_4$ detector
running during three years, and could hence be very competitive with existing experiments. Moreover, using this unique kind of direct search of Dark Matter
 could lead to unambiguous arguments in favor of a positive detection of Dark Matter. Also, we have shown that a directional detection of Dark Matter will allow 
 to constrain the Dark Matter properties both from particle physics and galactic halo physics. However, directional detection represents a major experimental challenge. In
 this context, the MIMAC experiment has been proposed and combined with a dedicated likelihood analysis, we have shown that MIMAC should be able to perform a 
 highly performant directional  detection of Dark Matter.

\end{document}